\documentclass[aps,prb,floatfix,twocolumn,footinbib]{revtex4-1} 
\usepackage{epsfig}
\usepackage{hyperref}
\usepackage{cancel} 
\usepackage{amsmath, amssymb} 
\usepackage{graphicx}
\usepackage{epsf}
\usepackage{braket}
\usepackage{appendix}
\usepackage{color}

\graphicspath{{Figures/}}

\begin{document}

\title{Temperature-tunable semiconductor metamaterial}

\author{K.\,L. Koshelev$^{1,2}$}
\email{ki.koshelev@gmail.com}
\author{A.\,A. Bogdanov$^{1,2,3}$}
\email{bogdanov@ioffe.mail.ru}

\affiliation{
$^{1}$    ITMO University, 197101 St.~Petersburg, Russian Federation \\
$^{2}$    Ioffe Institute, 194021 St.~Petersburg, Russian Federation \\
$^{3}$    The Academic University, 194021 St.~Petersburg, Russian Federation}

\date{\today}

\begin{abstract}
 We propose a novel class of temperature-tunable semiconductor metamaterials that exhibit negative refraction in the terahertz spectral range. These metamaterials are based on doped semiconductor superlattices with ultrathin barriers of about 1 nm thickness. Due to the tunnel transparency of the barriers, layers of the superlattice cannot be considered as isolated and, therefore, the classical homogenization approach is inapplicable. We develop a theory of quantum homogenization which is based on the Kubo formula for conductivity. The proposed approach  takes into account the wave functions of the carriers, their distribution function and energy spectrum.  We show that the components of the dielectric tensor  of the semiconductor metamaterial can be efficiently manipulated by external temperature and a topological transition from the dielectric to hyperbolic regime of metamaterial can be observed at room temperature. Using a GaAs/Al$_{0.3}$Ga$_{0.7}$As superlattice slab as an example, we provide a numerical simulation of an experiment which shows that the topological transition can be observed in the reflection spectrum from the slab. 

\end{abstract}

\keywords{metamaterial, terahertz, tunable, semiconductor, hyperbolic}

\maketitle

\section{Introduction}

Hyperbolic metamaterials (HMMs) are one of the fastest developing branches of modern optics.~\cite{Guo2012,cortes2012quantum,noginov2013focus,Poddubny2013hyperbolic,Ferrari2015} The dielectric function of HMMs is described by a tensor with two different components corresponding to the directions along ($\varepsilon_\|$) and across ($\varepsilon_\bot$) the optical axis:
\begin{equation}
\label{eps_tensor}
   \varepsilon=
   \left(
   \begin{matrix} 
      \varepsilon_{\bot} & 0 & 0\\
      0 & \varepsilon_{\bot} & 0\\
      0 & 0 & \varepsilon_{||} \\
   \end{matrix}
   \right). 
\end{equation}
Depending on the sign of these components, the crystal represents a dielectric medium ($\varepsilon_\bot>0,\ \varepsilon_\|>0$), a metal ($\varepsilon_\bot<0,\ \varepsilon_\|<0$) or a hyperbolic metamaterial  ($\varepsilon_{\bot}\varepsilon_{\|}<0$). For HMMs the shape of equal-frequency surface in {\bf k}-space represents a one- or two-sheet hyperboloid depending on the signature of permittivity tensor.\cite{Poddubny2013hyperbolic} This results in a singularity of the photon density of states and explains the unique optical properties of HMMs.\cite{jacob2012broadband} 

In tunable metamaterials,  the dielectric and magnetic responses and, therefore, the shape of equal-frequency surface can be manipulated by external influences such as, for example, DC magnetic field,\cite{Ramovic2011,smolyaninov2014quantum,li2012switchable} temperature,\cite{smolyaninov2010metric,ou2011reconfigurable,zhu2011thermal} femtosecond light pulses,~\cite{macdonald2001light, Chen2008,Averitt2007} application of a voltage,~\cite{Chen2006,Benz2014,gabbay2012theory} illumination,\cite{Padilla2006,Degiron2007} etc. In highly tunable metamaterials, the signature of the permittivity tensor can be switched, changing the equal-frequency surface topology. This phenomenon is called topological transition. \cite{krishnamoorthy2012topological} It has been experimentally observed is some systems.\cite{smolyaninov2010metric,smolyaninov2014quantum}

Fabrication of tunable THz metamaterials is an important problem because of numerous potential applications in far-field subwavelength imaging,\cite{jacob2006optical} enhanced nonlinearities,\cite{wurtz2011designed} nanoscale wave guiding and strong light confinement.\cite{govyadinov2006metamaterial} Some realizations of tunable THz metamaterials were presented, for example, in Refs.~\onlinecite{tao2011mems} and \onlinecite{yao2011three}. The majority of them represent an array of resonators whose capacity and/or inductivity is changed by external influences.\cite{tao2011mems} 

Here we propose a new concept of an ultra homogeneous temperature tunable metamaterial based on a semiconductor superlattice for THz applications. Here, the term \textit{ultra homogeneous} implies that the superlattice consists of coupled quantum wells separated by thin ($\sim$ 1~nm) tunnel-transparent barriers. Superlattices with barriers of such a thickness are widely used, for example, in quantum cascade lasers.\cite{Gmachl2001}  In this case, in contrast to a superlattice with thick barriers, \cite{Plumridge2008,Shekhar2014} quantum effects are particularly relevant and, therefore, it is incorrect to describe the dielectric function of each layer separately and then apply the homogenization procedure. Therefore, another approach, which takes into account the wave functions of the carriers, their energy spectrum modified by the superlattice potential and the carrier distribution function, should be used. We discuss the theory of proper approximation (quantum homogenization) further in Sec.~II.

It has been shown that highly doped semiconductor superlattice can exhibit properties of a hyperbolic medium at infrared frequencies.\cite{Yang2012,naik2012demonstration,Shekhar2014,hoffman2007negative} The frequency range of the hyperbolic regime are defined by plasma frequency, which depends on the free carrier concentration. Free carrier concentration in semiconductors is extremely sensitive to the temperature, in contrast to dielectrics and metals. For example, in the vicinity of the donor activation temperature, it can change by several orders.\cite{sze2006physics} We have shown that high temperature sensitivity of plasma frequency in a semiconductor metamaterial can be exploited for the efficient tuning of the metamaterial's optical properties in the THz region and stimulation of the topological transition from the dielectric to hyperbolic regime.      

 \begin{figure*}[t]
\begin{minipage}{0.98\linewidth}
\center{\includegraphics[width=1\linewidth]{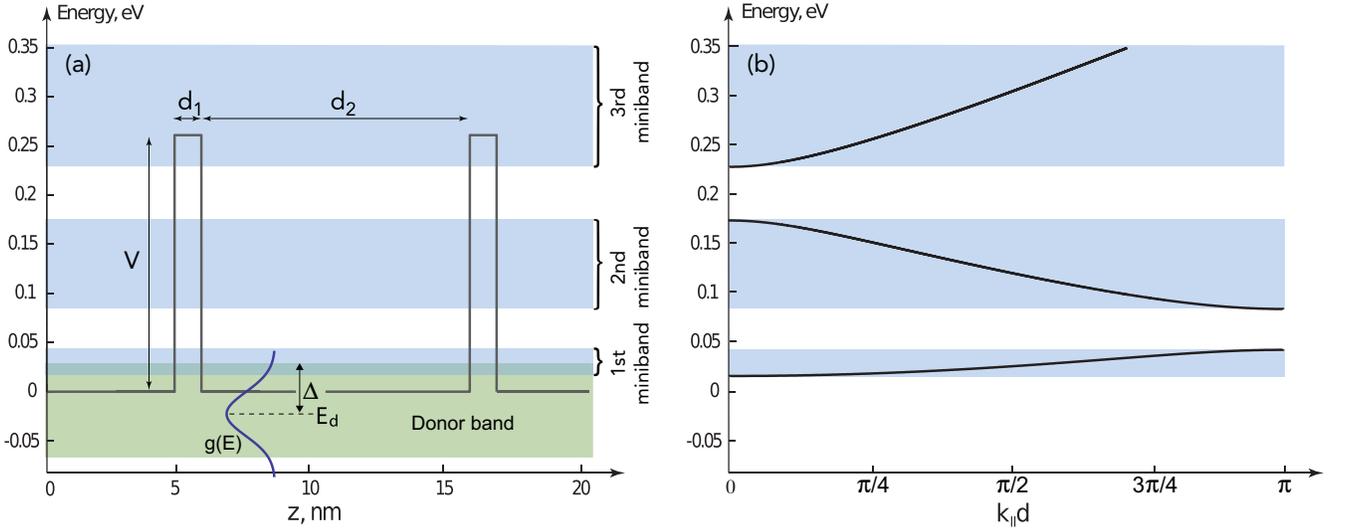}}
\caption{(a) Conduction band profile of $n$-doped superlattice. Blue shaded areas represent the energy of the minibands. Green shaded area corresponds to the donor band. Thicknesses of the well and the barrier are denoted as $d_1$ and $d_2$, respectively,  $V$ is the height of the barrier, $g(E)$ is a donor distribution function with standard deviation $\Delta$ and a maximum at $E_d$. (b) Electron energy dispersion in GaAs/Al$_{0.3}$Ga$_{0.7}$As superlattice with following parameters: $d_1$ = 10~mn, $d_2$ = 1~mn, $V$=~0.26~eV.}
\label{fig:1}
\end{minipage}
\end{figure*}

The paper is organized as follows. In Sec.~II  we develop a quantum homogenization theory and derive the main equations for the effective permittivity tensor. In Secs.~III and IV, we analyze the band structure and effective dielectric function of a Te-doped  GaAs/Al$_{0.3}$Ga$_{0.7}$As superlattice depending on the temperature and frequency of the electromagnetic field.  In Sec.~V, numerical simulation of an experiment on the measurement of the reflection spectrum from the superlattice slab is performed. In Sec.~VI we discuss a figure of merit and tunability of semiconductor metamaterials. Finally, in Sec.~VII we summarize our major results.

\section{Model}

\subsection{Quantum homogenization}

Within the effective medium approximation,  a multilayered structure with layer permittivities $\varepsilon_i$ and layer thicknesses $d_i$ can be considered as uniaxial optical crystal with permittivity tensor~(\ref{eps_tensor}) whose principle components are determined as
\begin{equation}
\frac{1}{\varepsilon_\bot}=\frac{1}{d}\sum\limits_i\frac{d_i}{\varepsilon_i},\ \ \ \ \varepsilon_\|=\frac{1}{d}\sum\limits_i d_i \varepsilon_i,\ \ \ \  d=\sum\limits_{i}d_i.
\end{equation}
As we have mentioned in the introduction, these formulas are inapplicable when the thickness of the layers is comparable with electron wavelength and, therefore, quantum mechanics laws become relevant. We consider a more accurate approach based on the Kubo formula.\cite{murayama2008mesoscopic} It takes into account the distribution function of the carriers, their wave functions and spectrum modified by the superlattice potential:   

\begin{equation}
\label{eq_Kubo}
\varepsilon_\alpha(\omega)=\varepsilon^{\infty}_\alpha\left(1-\frac{\Omega_\alpha^2}{\omega(\omega+i\gamma)}\right)+\frac{4\pi i}{\omega} \sigma_\alpha(\omega).
\end{equation}
Here and in what follows, the index $\alpha=\|,\bot$ corresponds to the directions along and across the optical axis of the metamaterial. Parameter $\varepsilon^\infty_\alpha$ is a permittivity of the lattice without free carriers, $\gamma$ is inverse momentum relaxation time of the carriers which is supposed to be isotropic for simplicity.

 The first term  interprets classical Drude-Lorentz formula. One can see that implementation of a superlattice in a semiconductor makes its plasma frequency anisotropic and we can distinguish plasma frequencies along ($\Omega_\|$) and across ($\Omega_\bot$) the optical axis:

 \begin{equation}
 \label{Plasma_freq}
\Omega_{\alpha}^2=\frac{4\pi e^2}{\varepsilon^{\infty}} \frac{2}{(2\pi\hbar)^3}\sum\limits_i\iiint f(E,\mu,T) \frac{\partial^2 E_i}{\partial p_{\alpha}^2}d^3p.
 \end{equation}
 Here $E_{i}$ is the carrier energy in the $i$-th miniband which depends on the momentum ${\bf p}$, $f(E,\mu,T)$ is the Fermi-Dirac distribution function, $\mu$ is the chemical potential, $T$ is the temperature. The sum is over all the minibands. Here we neglect hole contribution into the plasma frequency because we will consider $n$-doped semiconductor structures. Equation~(\ref{Plasma_freq}) is similar to the classical definition of plasma frequency: 
 \begin{equation}
 \label{Plasma_freq_class}
 \Omega^2=\frac{4\pi n e^2}{\varepsilon^\infty m^*}.
 \end{equation} 
Indeed, the difference between Eq.~(\ref{Plasma_freq}) and Eq.~(\ref{Plasma_freq_class}) is that in Eq.~(\ref{Plasma_freq}) we just average the inverse effective anisotropic mass $1/m^*=\partial^2{E}/\partial{p}^2$ with distribution function $f(E,\mu,T)$.

\begin{figure*}[t]
\begin{minipage}{0.98\linewidth}
\center{\includegraphics[width=1\linewidth]{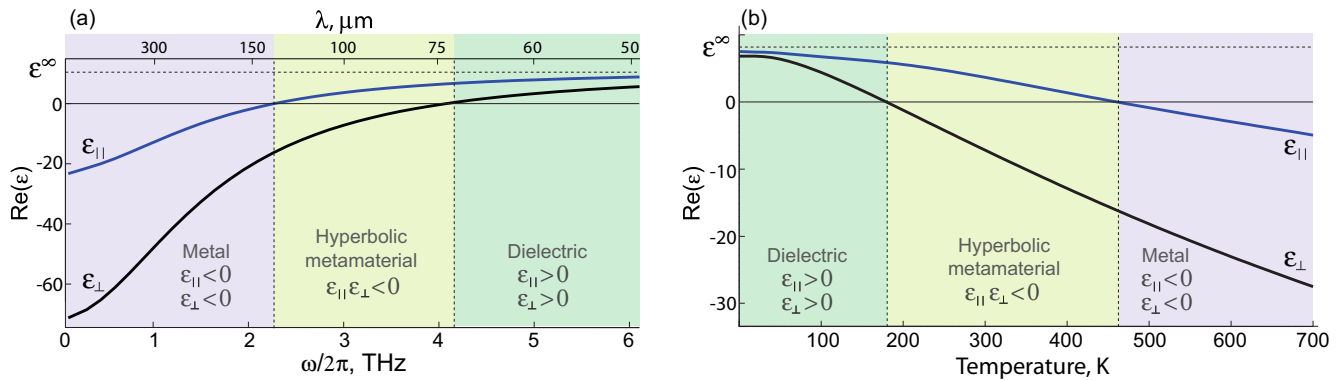} }
\caption{Color online. (a) Frequency dependence of real part of dielectric function along (blue line) and across (black line) the optical axe. Temperature T~=~300~K. (b)  Temperature dependence of real part of dielectric function along (blue line) and across (black line) the optical axis. Permittivity of the media without free carriers $\varepsilon^\infty$ is shown by the dashed line. Frequency $\omega/2\pi$ = 3~THz.}
\label{fig:2}
\end{minipage}
\end{figure*}

Second term in Eq.~(\ref{eq_Kubo}) describes interband transitions:
\begin{equation}
\sigma_\alpha=\frac{2i}{(2\pi\hbar)^3}\sum_{i,j}\iiint\frac{f(E_i)-f(E_j)}{\omega_{ij}-\omega+i\gamma}
\frac{|\hat J_{\alpha}^{ij}|^2}{\hbar\omega_{ji}} d^3p
\end{equation}

Here $\hat J_{\alpha}^{ij}=\bra{i,\mathbf{p}}\hat J_{\alpha}\ket{j, \mathbf{p}}$ is the matrix element of the current operator $\hat J_\alpha$ between the unperturbed eigenstates of the superlattice and  $\hbar\omega_{ij}=E_i-E_j$.

In order to calculate $\Omega_\bot$ and $\Omega_\|$  we need to determine the energy spectrum $E_i(\mathbf p)$ and the chemical potential $\mu$.

\subsection{Energy spectrum of carriers}

Let us consider a periodic semiconductor superlattice with period $d$ consisting of a quantum well with thickness $d_1$ and a barrier with thickness $d_2$ and height $V$ [Fig.~\ref{fig:1}(a)]. Effective masses in the well and barrier we put equal to $m_1$ and $m_2$, respectively. 

The total energy of an electron in a superlattice can be represented as the sum of two terms which correspond to electron motion across and along the optical axis:

\begin{equation}
    E_i(\mathbf{p})=\dfrac{p_\bot^2}{2m_\bot^*}+E_i(p_\|)
\end{equation}
Here $m_\bot$ is the effective mass of electrons in the direction across the optical axis. We take it equal to the effective electron mass in a bulk material. The second term of the right-hand side describes energy dispersion of electrons in $i$-th miniband. It can be found from the dispersion equation.
    \begin{multline}
    \label{eq1}
    \cos({p_\| d/\hbar})=\cos({p_1 d_1/\hbar})\cos({p_2 d_2/\hbar}) -\\
    \\-\frac{1}{2}\sin({p_1 d_1/\hbar})\sin({p_2 d_2/\hbar})\left(\frac{p_1m_2}{p_2m_1}+\frac{p_2m_1}{p_1m_2}\right)\\
    \end{multline}
    where $ p_1=\sqrt{2m_1 E(p_\|)},\  p_2=\sqrt{2m_2(E(p_\|)-V)}$. This dispersion equation can be obtained from the Schr$\ddot{\text{o}}$dinger equation using the Floquet's theorem.

\subsection{Chemical potential}

Efficient temperature manipulation of free carrier concentration can be realized in narrow-gap or in doped semiconductor structures. Temperature tuning in metamaterials based on narrow-gap semiconductors was partially analyzed in Ref.~\onlinecite{bui2013thermally}.   Here we consider the case of doped semiconductor structures on the example of a superlattice with quantum wells uniformly doped with shallow donors.

In highly doped structures, wave functions of neighbour donors can overlap. This results in a shift of donor levels and formation of a donor band. In the case of a considerable shift, the donor band can overlap with the conduction band. This phenomenon is called the Mott transition and will be discussed further.    

The chemical potential $\mu$ can be calculated from the electroneutrality condition~\cite{sze2006physics}. It states that free carrier concentration is equal to the concentration of ionized donors:

\begin{equation}
\label{eq6}
\sum_{i}\int{f(\textbf{p},\mu,T)}\frac{2d^3p}{(2\pi\hbar)^3}=n_d \int\frac{ g(E)dE}{2e^{(\mu-E)/(kT)}+1}
\end{equation}
Here $n_d$ is the full donor concentration, $g(E)$ is a donor distribution function which can be approximated by a Gaussian~\footnote{ For the sake of simplicity, we neglect effects of the Coulomb gap~\cite{shklovskii1979electronic,efros1975coulomb}} with standard deviation $\Delta$ and maximum at $E_d$.\cite{brody1962nature}

\section{Band structure of the superlattice \label{subsec:band_structure}}

The model described above is applicable for superlattices of various compounds and designs. As an example, let us consider a superlattice that consists of GaAs quantum wells and  Al$_{0.3}$Ga$_{0.7}$As barriers. Thickness of the quantum well $d_1$ and the barrier $d_2$ we put equal to 10 and 1 nm, respectively. We consider the high frequency permittivity in Eq.~(\ref{eq_Kubo}) to be isotropic ($\varepsilon_\|^\infty=\varepsilon_\bot^\infty$) and put it equal to 11.\cite{adachi1994gaas}  The calculated band structure and electron dispersion are shown in Fig.~\ref{fig:1}(b). We consider that quantum wells are uniformly doped with Te donors with concentration $n_d$ = 1 $\times$ 10$^{18}$~cm$^{-3}$. The energies of Te donors in bulk GaAs and in the superlattice are slightly different due to quantum confinement that arises from the superlattice potential. We neglect this difference and put $E_d$~=~0.03~eV as in a bulk material.\cite{sze2006physics} The standard deviation of the donor distribution function $\Delta$ for such a doping level is about several hundredths of an electron-volt. In the structure under consideration we put $\Delta$ = 0.03~eV, which is in accordance with Refs.~\onlinecite{brody1962nature} and \onlinecite{conwell1956impurity}.

\begin{figure}[t]
\begin{minipage}[t]{0.98\linewidth}
\center{\includegraphics[width=0.98\linewidth]{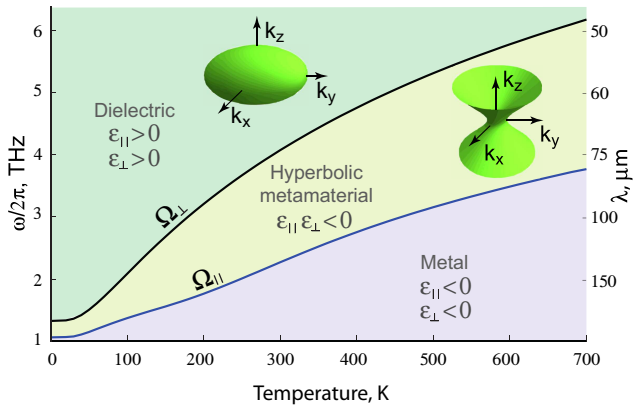} }
\caption{Color online. Temperature dependence of plasma frequency along the optical axis (blue line) and across the optical axis (black line). Insets show the shape of equal-frequency surfaces in the dielectric and hyperbolic regimes.}
\label{fig:4}
\end{minipage}
\end{figure}

The width of the first miniband is about several hundredths of an electron-volt, which is comparable with the gap between the minibands. So, both weak-coupling and tight-binding approximations are poorly applicable for the structure under consideration. 

For the chosen parameters, the overlap of donor miniband and first miniband is approximately equal to 0.01~eV. This means that even at zero temperature there are free carriers in the conduction band and their Fermi's energy is about  0.01~eV. This corresponds to a concentration about $10^{16}$~cm$^{-3}$. A simple estimation of plasma frequency using Eq.~(\ref{Plasma_freq_class}) yields $\Omega/(2\pi)\sim1$~THz.

\section{Effective dielectric function of the superlattice}

We calculate the frequency and temperature dependencies of the permittivity tensor components using the quantum homogenization approach  [Eq.~(\ref{Plasma_freq})]. The frequency dependence of $\varepsilon_{\bot}$ and $\varepsilon_{||}$ at room temperature is shown in Fig.~\ref{fig:2}(a). The average energy between minibands and, therefore, frequencies of interband transitions are about ~0.05 eV [Fig.~\ref{fig:1}(a)], which corresponds to a frequency of 10~THz. So, the contribution of interband transition into the dielectric function [second term in Eq.~(\ref{eq_Kubo})] can be neglected at frequencies of 1~THz  without any considerable precision losses. Thus, quantum homogenization predicts that, beyond the interband transition, the tensor components of a superlattice with thin layers can be described within the classical Drude-Lorentz formula with plasma frequency described by Eq.~(\ref{Plasma_freq}). 

\begin{figure}[t]
\begin{minipage}[t]{0.98\linewidth}
\center{\includegraphics[width=1\linewidth]{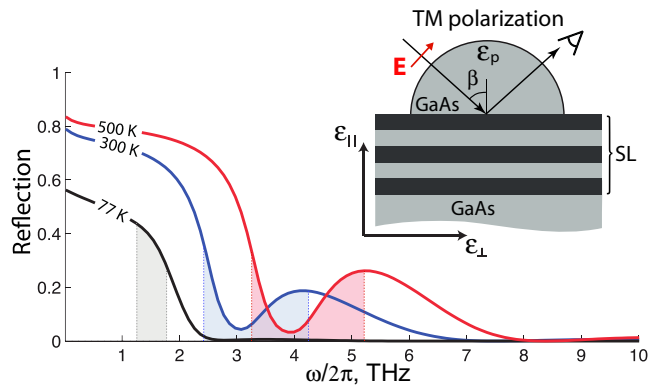} }
\caption{Color online. Frequency dependence of reflectance at three different temperatures. Shaded areas correspond to frequency bands of hyperbolic dispersion. The inset shows the relative orientation of the dielectric function ($\varepsilon_\bot$ and $\varepsilon_\|$), the electric-field vector (transverse magnetic), the layered structure and a spherical prism with permittivity $\varepsilon_p$ = 11. Angle of incidence is $\beta$ = 40$^0$.}
\label{fig:5}
\end{minipage}
\end{figure}

This results qualitatively differ from effective parameters obtained within the classical homogenization procedure which predicts a resonance behaviour of $\varepsilon_\|$ at a nonzero frequency. There is no contradiction here. Classical homogenization implies that all layers are isolated from each other and that the carriers do not move from one layer into a neighbouring one. Qualitatively it is equivalent to the restoring force that obstructs the carrier transport. This force results in an appearance of the resonance in $\varepsilon_\|$.  Quantum homogenization implies that the barriers are tunnel transparent and charges can move freely throughout the whole volume of the sample. Therefore, the dielectric function $\varepsilon_\|$ is similar to that of a metal but takes into account interband transitions.                   

One can see from Fig.~\ref{fig:2}(a) that there are three frequency regions which correspond to different forms of the equal-frequency surfaces in k-space: (i) at frequencies $\omega/2\pi>4.1$~THz the material behaves like a dielectric; (ii) at  frequencies $2.2$~THz $<\omega/2\pi<4.1$~THz the form  of the equal-frequency surface is a hyperboloid and the material exhibits optical properties  of HMM; (iii) at $\omega/2\pi<2.2$~THz electromagnetic waves decay exponentially into the medium, similar to the behaviour in a metal.

     The temperature dependence of $\varepsilon_{\bot}$ and $\varepsilon_{||}$ at a frequency of 3~THz  is shown in Fig.~\ref{fig:2}(b). One can see that the material behaves as a dielectric, HMM or a metal depending on the temperature. Dielectric dispersion can be realized at the temperature of liquid nitrogen, the hyperbolic regime is achieved at room temperature. Thus, a topological transition for THz radiation can be realized at temperatures reasonable for an experiment. 
     
     The temperature dependence of the permittivity tensor components is weak at low temperatures. This is explained by the fact that the main part of donor levels are noticeably separated from the conduction band bottom, thus the activation of electrons on such levels occurs at temperature $kT\sim E_D$. At higher temperatures when all donors are ionized, the dielectric functions $\varepsilon_{\bot}$ and $\varepsilon_{||}$ tend to constant values. In our case it occurs at temperatures  much higher than the melting point of GaAs which is 1511~K. \cite{adachi1994gaas}
     
Figure~\ref{fig:4} represents a topological phase state diagram. Solid lines show the temperature dependence of the longitudinal $\Omega_\|$ and transversal $\Omega_\bot$ plasma frequencies. These lines divide the plane of the figure into three regions. It follows from Eq.~(\ref{eq_Kubo}) that every region corresponds to the one of the possible regimes: dielectric, metal or hyperbolic. The shapes of equal-frequency surfaces corresponding to each regimes are shown in the insets of Fig.~\ref{fig:4}. 

One can see that plasma frequencies $\Omega_\|$ and $\Omega_\bot$ are not equal to zero at low temperatures. This is explained by the overlapping of the donor and first minibands, the so-called Mott transition \cite{shklovskii1979electronic}. A part of the carriers does not freeze-out at low temperatures and makes a contribution into the plasma frequencies.

\section{Simulation of experiment}

\begin{figure}[t]
\begin{minipage}[t]{0.98\linewidth}
\center{\includegraphics[width=1\linewidth]{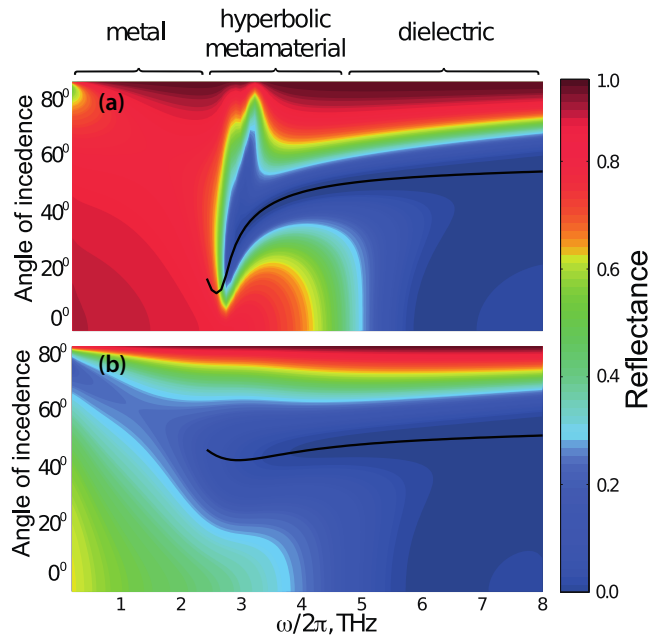} }
\caption{Color online.  Simulation of reflectance spectrum from 10~$\mu$m GaAs/Al$_{0.3}$Ga$_{0.7}$As superlattice for different incident angles $\beta$. Incident wave has TM-polarization. Parameters of the superlattice is mentioned in Sec.~\ref{subsec:band_structure}. Temperature is taken equal to 300~K. Inverse momentum relaxation time is: (a)  $\gamma$ = 1 $\times$ 10$^{12}$~s$^{-1}$; (b) $\gamma$= 1 $\times$ 10$^{13}$~s$^{-1}$. Black solid line corresponds to the Brewster's angle which is determined by Eq.~(\ref{brewster_angle}).}
\label{fig:6}
\end{minipage}
\end{figure}
The efficiency of temperature tuning and the presence of a topological transition in a metamaterial based on a semiconductor superlattice can be confirmed experimentally by the measurement of reflection spectrum from the metamaterial. Here we provide a numerical simulation of a possible experiment.The scheme of the experiment is shown in the inset of Fig.~\ref{fig:5}. A plane electromagnetic wave of TM-polarization is supposed to be incident at the angle $\beta$ on the 10~$\mu$m GaAs/Al$_{0.3}$Ga$_{0.7}$ superlattice grown on an undoped GaAs substrate. The inverse momentum relaxation time $\gamma$ in Eq.~(\ref{eq_Kubo}) we put equal to  3~$\times$~10$^{12}$~s$^{-1}$. For the sake of simplicity, we neglect temperature dependence of $\gamma$. In order to get efficient excitation of the optical states with high wave vectors, we consider that the wave incident on the sample passes through an undoped GaAs spherical prism with permittivity $\varepsilon_p=11$.

The frequency dependence of the reflectance for different temperatures and a fixed incident angle $\beta$~=~40$^0$ is shown in Fig.~\ref{fig:5}. Frequency intervals of the hyperbolic regime are marked by shadows. For the temperatures 300~K and 500~K there is a minimum of reflection in the hyperbolic regime. The nature of the minimum can be explained in terms of Brewster's angle and will be discussed further. For the temperature 77~K the minimum is smeared out because the frequency width of hyperbolic regime is comparable with Drude relaxation constant $\gamma$.        

The frequency interval corresponding to the hyperbolic regime strongly depends on the temperature. So, at 77~K the metamaterial exhibits properties of HMM at frequencies $1.2$~THz~$<\omega/2\pi<1.7$~THz. At room temperature this frequency interval is $2.5$~THz $<\omega/2\pi<4.2$~THz. Temperature drastically affects the reflection coefficient. For example, increasing of the temperature from 77~K to 300~K at the frequency $2.5$~THz results in tenfold increase of reflection coefficient from 0.03 to 0.3.  

Figure~\ref{fig:6} shows the reflectance at room temperature versus incident angle and wave frequency. Typical values of inverse momentum relaxation time $\gamma$ for semiconductors lie in the interval between 10$^{12}$~s$^{-1}$ and 10$^{13}$~s$^{-1}$ (e.g., see Refs.~\onlinecite{sze2006physics} and \onlinecite{law2014doped}). Subfigures (a) and (b) correspond to these limit cases. One can see from Fig.~\ref{fig:6}(a) that in the case of low losses,  dielectric, metal, and hyperbolic regimes have well-distinct frequency boundaries which smear out with increasing of the losses [Fig.~\ref{fig:6}(b)].

In the dielectric regime ($\omega/2\pi\gtrsim$~4.0~THz) we have total reflection for glancing incident angles  ($\beta\gtrsim80^0$). Total reflection takes place because $\varepsilon_p>\varepsilon_\|>\varepsilon_\bot>0$. At incident angles less than $80^0$ ($\beta< 80^0$) we have nearly total transmission and a minimum of the reflection coefficient in the vicinity of $\beta$~=~50$^0$. High transmission is explained by low optical contrast between the sample and prism in the dielectric regime. The minimum corresponds to the Brewster's angle $\alpha_B$. For an uniaxial crystal $\alpha_B$ can be determined as\cite{shu2007brewster}
\begin{equation}
\alpha_B=\text{Re}\left\{\arcsin\left(\sqrt{\frac{\varepsilon_\bot\varepsilon_\|-\varepsilon_\|\varepsilon_p}{\varepsilon_\bot\varepsilon_\|-\varepsilon_p^2}}\right)\right\}.
\label{brewster_angle}
\end{equation}
Frequency dependence of Brewster's angle $\alpha_B$ is shown in Figs.~\ref{fig:6}(a) and (b) by black solid line. 

In the metal regime ($\omega/2\pi<$~2.2~THz) we have nearly total reflection for all angles of incidence except the minimum in the vicinity of $\beta~=$ 85$^0$ at low frequencies. The minimum corresponds to the excitation of the surface plasmon polariton mode in the slab.

In the hyperbolic regime (2.2~THz~$<\omega/2\pi<$~4.0~THz), there are regions of high and low reflection. The high reflection region appears because the HMM under consideration does not support optical states with small lateral (lying in plane of the interface) components of the wave vector. Low reflectance in the region for $\beta$ from 20$^0$ to 50$^0$ can be explained by the Brewster's angle when the TM polarized wave is not reflected from the sample. The minimum of reflection at $\beta~\sim$~60$^0$ is explained by the excitation of the waveguide mode in HMM slab.\cite{Bogdanov2011,Bogdanov2012}

\section{FoM vs tunability}

The optimal operation frequency for semiconductor tunable metamaterials can be estimated from the condition of high tunability and low losses. The main absorption mechanisms in semiconductors are free carrier absorption, fundamental absorption, and absorption due to the optical phonons. \cite{sze2006physics} Fundamental absorption is dominant in the optical range. Absorption due to optical phonons plays an important role in polar semiconductors. It has resonance behaviour and is significant around the wavelengths 30-35~$\mu m$. Free carrier absorption increases with wavelength as square of the wavelength, so these losses make the main contribution in the THz region. Rough estimation using Eq.~(\ref{eq_Kubo}) shows that figure of merit (FOM) $\eta$ for $\omega/2\pi$=3~THz and $\gamma$ = 1~$\times$~10$^{13}$~s$^{-1}$ is

\begin{equation}
\label{FoM_def}
\eta=\left|\frac{\mbox{Re}(\varepsilon)}{\mbox{Im}(\varepsilon)}\right|=\frac{\omega}{\gamma}\approx2.
\end{equation}
Therefore, high FOM is achievable for $\omega\gg\gamma$. 

From the other hand, high tunability of optical properties is reached near or below the plasma frequency [see Eq.~(\ref{eq_Kubo})]:
\begin{equation}
\label{Tune_condition}
\omega\lesssim\Omega_p.
\end{equation}
Typical effective electron mass $m^*$ in semiconductors varies from $\sim0.01m_e$ for narrow-band semiconductors (InAs, InSb) to ~$\sim0.1m_e$ (Si, GaAs), permittivity $\varepsilon^\infty\sim10$, electron concentration varies in a wide range from vanishingly small values to $\sim$ 10$^{20}$~cm$^{-3}$ for highly doped semiconductors.\cite{law2014doped} Simple estimation using Eq.~(\ref{Plasma_freq_class}) yields the value of plasma frequency $\Omega_p/2\pi\sim$ 1~$\times$~10$^{14}$~Hz for free carrier concentration 1 $\times$ 10$^{20}$~cm$^{-3}$ and $m^*\sim0.1m_e$. It corresponds to the wavelength $\lambda_p\sim$ 3~$\mu$m. Therefore, semiconductors can potentially be used for the fabrication of tunable metamaterials for near infrared applications. However it is very difficult to reach effective tunability for highly doped semiconductors because of the considerable screening and strong random potential.\cite{sze2006physics,shklovskii1979electronic} The simplest way to avoid these problems is to decrease the doping level to   10$^{17}$-10$^{18}$~cm$^{-3}$. Such concentrations correspond to plasma frequencies $\Omega/2\pi\sim$ 0.3 - 1~$\times$~10$^{13}$~Hz, i.e. to the wavelength $\lambda_p\sim$ 30 - 100~$\mu$m.

Therefore, effective tunability and low losses in semiconductor metamaterials can be reached in the frequency range $\gamma\ll\omega\lesssim\Omega_p$ which corresponds to THz frequency band.

\section{Conclusion}
In this work we proposed a new concept of an ultra homogeneous temperature-tunable metamaterial based on a doped semiconductor superlattice. We have shown that the classical homogenization procedure is inapplicable for the description of the metamaterial in terms of effective parameters because of the tunnel transparency of the barriers separating the quantum wells and that quantum homogenization should be used. We developed the theory of quantum homogenization applied to semiconductor nanostructured metamaterials. It is based on the Kubo formula for conductivity and takes into account wave functions of the carriers, their energy spectrum and distribution function. 

We have shown that the components of the dielectric tensor  the semiconductor metamaterial can be efficiently manipulated by external temperature. Efficient temperature tunability is a distinctive feature of semiconductors which is explained by the high sensitivity of free carrier concentration to the temperature. On the example of a GaAs/Al$_{0.3}$Ga$_{0.7}$As semiconductor superlattice with Te-doped quantum wells ($n_d$ = 1 $\times$ 10$^{18}$~cm$^{-3}$)  we have shown that the temperature of the topological transition from the dielectric to hyperbolic regime $\sim$ 300~K for the frequency $\sim$ 4~THz. Numerical simulation shows that the topological transition can be detected in an experiment on the measurement of reflection spectrum from the superlattice slab. 

A significant advantage of semiconductor metamaterials is the possibility of their direct integration into optolectronic devices and optical integrated circuits.   Moreover, semiconductor materials combine two important features. On one hand, the energy spectrum of the carriers in semiconductor nanostructures can be precisely tailored with quantum engineering technologies. On the other hand, there are many methods of dynamic control of the electron distribution function in semiconductors, which are well-developed and widely applied in nano- and optoelectronics. These are, for example, electrical injection, optical pumping, thermal excitation, electron heating by electric field, etc. The advantages mentioned earlier and the rich functionality of semiconductor metamaterials allow to consider them as important element of future optoelectronics.

\acknowledgements

This work was partially supported by the Government of the Russian Federation (Grant 074-U01), by the Russian Foundation for Basic Research and by the Program on Fundamental Research in Nanotechnology and Nanomaterials of the Presidium of the Russian Academy of Sciences.


 \bibliography{main}

 \end{document}